# Galactic Outflows and the pollution of the Galactic Environment by Supernovae[*]

Elisabete M. de Gouveia Dal Pino (1), Claudio Melioli (1,2), Anibale D'Ercole (2), Fabrizzio Brighenti (3), Alex Raga (4)

(1) IAG-USP, Universidade de São Paulo, R. do Matão 1226, 05508-090 São Paulo, SP, Brazil
(2) Dipartimento di Astronomia, Università di Bologna, via Ranzani 1, 40126 Bologna, Italy
(3) INAF-Osservatorio Astronomico di Bologna, via Ranzani 1, 40126 Bologna, Italy
(4) ICN-UNAM, Universidad Autónoma de Mexico, DF, Mexico

**Abstract**

We here explore the effects of the SN explosions into the environment of star-forming galaxies like the Milky Way. Successive randomly distributed and clustered SNe explosions cause the formation of hot superbubbles that drive either fountains or galactic winds above the galactic disk, depending on the amount and concentration of energy that is injected by the SNe. In a galactic fountain, the ejected gas is re-captured by the gravitational potential and falls back onto the disk. From 3D non-equilibrium radiative cooling hydrodynamical simulations of these fountains, we find that they may reach altitudes up to about 5 kpc in the halo and thus allow for the formation of the so called intermediate-velocity-clouds (IVCs) which are often observed in the halos of disk galaxies. The high-velocity-clouds that are also observed but at higher altitudes (of up to 12 kpc) require another mechanism to explain their production. We argue that they could be formed either by the capture of gas from the intergalactic medium and/or by the action of magnetic fields that are carried to the halo with the gas in the fountains. Due to angular momentum losses to the halo, we find that the fountain material falls back to smaller radii and is not largely spread over the galactic disk. Instead, the SNe ejecta fall nearby the region where the fountain was produced, a result which is consistent with recent chemical models of the galaxy. The fall back material leads to the formation of new generations of molecular clouds and to supersonic turbulence feedback in the disk.

## 1. Introduction: a brief overview

Edge-on star forming disk galaxies often exhibit hot halos of ionized gas that may extend up to several kpc of height over the regular HI galactic disk. They are fed by ascending gas from the disk in structures that resemble chimneys and fountains. Observations indicate that the chimneys are generated by supernovae (SNe) explosions which blow superbubbles that expand and carve holes in the disk injecting high speed, metal enriched gas which forces its way out through relatively narrow channels with widths of 100 - 150 pc. They establish a connection between the thin disk and the halo feeding it with the hot disk gas that expands through the halo under buoyancy forces up to a maximum height and then returns to the disk bent by the disk gravity. The whole cycle is like a fountain - hence the name galactic fountain. Close-up evidence for chimneys is scant, but evidence for large chimneys is clear in external galaxies in the form of holes in the distribution of HI, often with some evidence for flows. In the Milky Way (MW), the evidence has mainly been in the form of fragments and vertical structures in the large scale maps of the interstellar medium. A multi-

---

wavelength survey of the halos of several star forming galaxies (e.g., Dettmar 2005; Dettmar & Soida 2005) have revealed a correlation of these halos with the rates of star formation and the energy input rates by SNe suggesting that gaseous halos are associated to star formation processes in the disk.

Other observed features that seem to be correlated to gas circulation in chimneys and galactic fountains are the so called intermediate and high-velocity-clouds (IVCs and HVCs, respectively). These are mainly neutral hydrogen (HI) clouds that can be as large as 100 pc, with masses of up to $10^4$ solar masses which are observed in the halo of the MW and other star forming galaxies at altitudes typically between 300 pc and 2.5 kpc and which are falling on the disk with velocities between -20 km/s e -90 km/s. The HVCs can be observed in even higher altitudes of up to 12 kpc and velocities up to -140 km/s. Figure 1 provides a mosaic of the matter distribution in the halo of the MW. It indicates that the galactic disk is surrounded by a very *cloudy* environment. It is generally believed that at least the IVCs have been formed from the condensation of the gas that arises in the chimneys triggered by SN explosions and numerical simulations suggest that this seems to be indeed the case (e.g., de Avillez 2000; de Avillez & Berry 2001; Melioli, Brighenti, D'Ercole, de Gouveia Dal Pino 2008 a, 2008b, see below). However, the origin of the HVCs is still controversial. The difficulty at producing fountains in the hydrodynamical simulations reaching altitudes higher than 5 kpc (Melioli et al. 2008 b) and the very small metallicity contents observed in these HVCs suggest that they may have been originated from gas raining into the galaxy that is accreted from the intergalactic medium (IGM) or from satellite galaxies (e.g., Fraternali & Binney 2006).

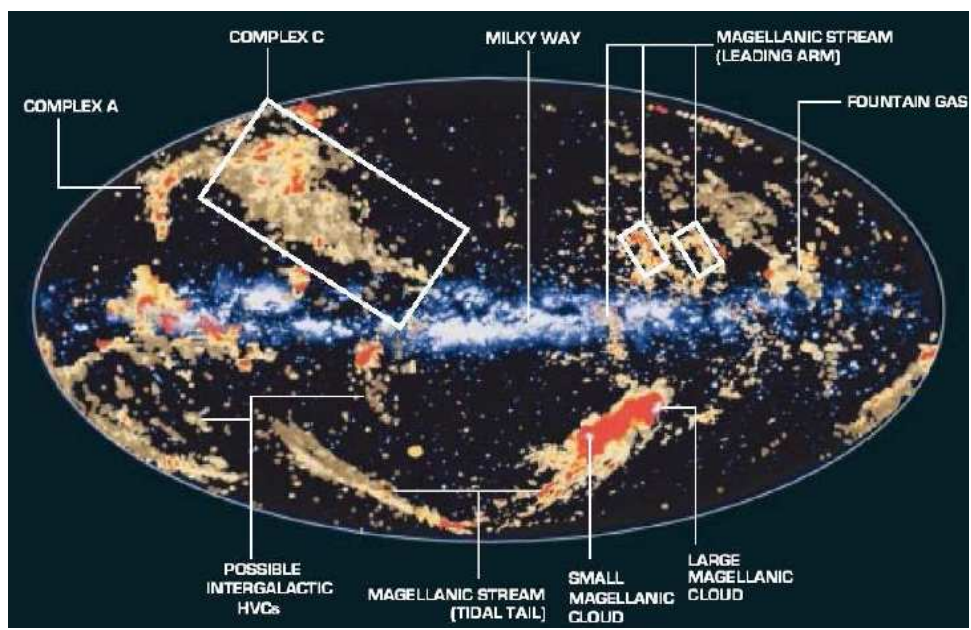

**Figure 1.** Map of the galactic gas and its environment: it combines radio observations of neutral hydrogen (HI) of the environment with a visible light image of the MW (the galactic disk in the middle). The high and intermediate-velocity clouds of hydrogen, such as complexes A and C, are located above and below the disk. A galactic fountain is also identified in the map (extracted from http://apod.nasa.gov).

Another extreme example of gas outflow from the disk of star forming galaxies are the supersonic winds which are powerful enough to escape from the gravitational potential of the galactic disk to the IGM. Spectacular winds extending for several kpcs above the disk have been observed in galaxies with bursts of star formation, the so called starburst (SB) galaxies. These are merging or interacting galaxies that may have star formation rates up to 20 times larger than those of regular galaxies, like the Milky Way, and this could explain the large amount of energy injection by SNe and the resulting production of powerful winds. Indeed, galactic winds are ubiquitous in SB galaxies (see Veilleux, Cecil & Bland-Hawthorn 2005 and references therein). Recent high resolution observations of the best studied prototype of this class, the Starburst galaxy M82, has evidenced that its wind, like in the galactic fountains, is being fed by SNe explosions nestled in several stellar associations around the nuclear region of the galaxy (Konstantopoulos et al. 2008). However, we have recently investigated the real effectiveness of the energizing process by the SNe in SBs and found that it is sensitive to the amount of gas locked within clouds. If large enough, then the efficient radiative cooling of the shock-heated clouds by the SNe may prevent the gas from escaping as a wind for about half-lifetime of the starburst (Melioli & de Gouveia Dal Pino 2004). In the present work, we are not going to address these powerful winds, but we refer to some models for their production (see e.g., Tomisaka & Ikeuchi 1988; Tomisaka & Bregman 1993; Suchkov et al. 1996; de Gouveia Dal Pino & Medina-Tanco 1999; D'Ercole & Brighenti 1999; Leitherer et al. 1999; Tenorio-Tagle & Munoz-Tunon 1998; Strickland & Stevens 2000; Tenorio-Tagle et al. 2003; Cooper, Bicknell &Sutherland 2008).

Extensive work on the formation of galactic chimneys and fountains has been carried out over the last decades (see e.g., de Avillez 2000; de Avilles & Breitschwerdt 2005; Melioli et al. 2008a, and references therein for reviews). Shapiro & Field (1976) first proposed the idea that galactic chimneys induced by SNe explosions would cause gas circulation between the disk and the halo. This scenario was afterwards explored analytically in detail by Bregman (1980) e Kahn (1981). More recently, the advent of powerful computers have made it possible to simulate fountains and winds rather accurately.

The first 3D hydrodynamical simulations following the whole cycle of the gas between the disk and the halo in fountains in the Milky Way were performed by de Avillez (2000) and de Avillez & Berry (2001). However, in order to obtain a high spatial resolution, they considered only a small region of the Galaxy with a dimension of 1 kpc$^2$ (in the disk) x 10 kpc (for the height in the halo). Korpi et al. (1999) have included the effects of the differential rotation and the magnetic field of the galactic disk, but considered a computational domain too small (500 pc$^2$ x 1 kpc) to allow for the development of an entire cycle of the chimney gas between the disk and the halo. Other works which were also concerned with the collective effects of supernovae on the structure of the interstellar medium (ISM) have considered even smaller volumes in their hydrodynamical simulations (see, e.g., MacLow, McCray & Norman, 1989). More recently, further MHD simulations were carried out by de Avillez & Breidtschwerdt (2005), without including differential rotation, where again, in order to reach very high resolution (of 0.6 pc) they considered only a small volume of the Galaxy (1 kpc$^2$ x 10 kpc), as they were primarily concerned at examining the role of the disk-halo gas circulation in establishing the volume filling factors of the different phases of the ISM in the Galactic disk. In particular, these MHD studies have revealed that the gas transport into the halo is not prevented by the parallel magnetic field of the Galaxy (as suggested in former works; see e.g., Tomisaka 1998), but only delayed by few tens of Myr when compared to pure HD simulations.

In this work, we summarize the results of recent study of the large scale development of galactic fountains driven by SNe explosions which was carried out in order to understand their dynamical evolution, the observed kinematics of the extraplanar gas, the IVCs and HVCs formation, and the influence of the fountains in the redistribution of the freshly delivered metals over the galactic disk. To this aim, we have performed fully 3D non-equilibrium radiative cooling hydrodynamical simulations of the gas in the Milky Way where the whole Galaxy structure, the galactic differential rotation and the supernovae explosions generated in single and multiple stellar associations of OB stars have been considered (see Melioli et al. 2008a, 2008b for more details).

## 2. Setup and Results for Galactic Fountains

The initial setup is described in detail in Melioli et al. (2008a, 2008b). The ISM is initially set in rotational equilibrium in the galactic gravitational potential well given by the summation of the dark-matter halo, the bulge and the disk contributions. Assuming the hydrostatic equilibrium in the z (vertical) direction between gravity and the full pressure (given by the thermal gas + magnetic + cosmic ray pressure contributions) we have computed the rotation velocity as a function of the galactic-centric distance R  This has allowed us to build the full galactic rotation curve, as observed in the MW for different z heights. The ISM in our model is made up of the three gas components, namely the molecular (H2), the neutral (H I) and the (H II) hydrogen and their stratified density distribution follows the empiric curves obtained for the MW by Wolfire et al. (2003). In most models a hot ($T_h = 7 \times 10^6$ K) isothermal gas halo has been added which is in equilibrium with the galactic potential well. In some models the halo has been allowed to rotate with a velocity that is a fraction of the disk velocity. In order to drive the formation of chimneys and fountains, we have considered two sets of models. In one set, we exploded continuously 100 SNe over 30 Myr in a single star cluster. In a second set, we have exploded randomly up 2000 SNe in multiple star clusters (spread over an area of either 1 kpc$^2$ or 8 kpc$^2$ on the disk). In both cases, the clusters were localized, in general, at a radial distance R= 8.5 kpc which corresponds to the Sun distance from the galactic centre. We have assumed the SNe rate distribution over time that has been inferred from observations of the MW (Higdon & Lingenfelter 2005).

In our simulations we have employed a modified version of the adaptive mesh refinement YGUAZU code that integrates the 3D invisced gasdynamic equations with the flux vector splitting algorithm of van Leer (Raga et al. 2000, 2001; see also, e.g., Gonzalez et al. 2004; Melioli, de Gouveia Dal Pino & Raga 2005). The non-equilibrium radiative cooling of the gas is computed together with a set of continuity equations for atomic/ionic or chemical species. The 3D binary, hierarchical computational grid is structured with a base grid, and with a number of higher resolution grids. We have enforced the maximum grid resolution only in the volume encompassed by the fountain. In order to follow the circulation and the thermal history (i.e. the degree of radiative cooling) of the metals expelled by the SNe, we have added three different tracers passively advected by the code describing the disk gas, the halo gas, and the SNe ejecta.(see Melioli et al. 2008a for details).

Figures 2 and 3 show the edge-on and the face-on view of the evolution of a galactic fountain arising from SNe explosions within a single OB association located at the galactic-centric distance R=8.5 kpc. At this radius the transition between the disk and the hot halo occurs at z = 800 pc. The critical luminosity for the SNe to break through the disk is $L_b = 1.5 \times 10^{37}$ erg/s (e.g., Koo & McKee 1992) well below the mechanical luminosity $L_W = 10^{38}$ erg/s provided by the 100 SNe powering the fountain. This model has been run with a maximum spatial resolution of 12.5 pc.

During its activity the fountain digs a hole on the disk and throws SNe ejecta and ISM vertically up to z~2 kpc above the galactic plane. Once the stellar explosions cease, the hole collapses in $2 \times 10^7$ yr and the ejecta trapped at its edges (nearly half of the total) mixes with the local ISM. Owing to the differential rotation of the Galactic disk, the ejecta does not remain confined in one point, but is stretched, giving rise to the bean-like structure seen in Fig. 2. The low density tail with a banana-shape is due to the ejecta pushed at high altitudes which then slowly falls back lasting above the disk. The two concentric circles in Fig. 2 with radii R = 8 kpc and R = 9 kpc, have been drawn to guide the eye. The SNe explode between these two circles, and their distance corresponds to the diameter of the hole produced by the fountain. Comparing the shape of the tail traced by the ejecta with the circles, we note that the gas of the fountain tends to move inward during its trajectory. Actually, as the gas moves upward and interacts with the halo, it transfers to it part of its angular momentum; the centrifugal force decreases in pace with the circular velocity and gravity prevails, pushing the gas toward the Galactic centre.

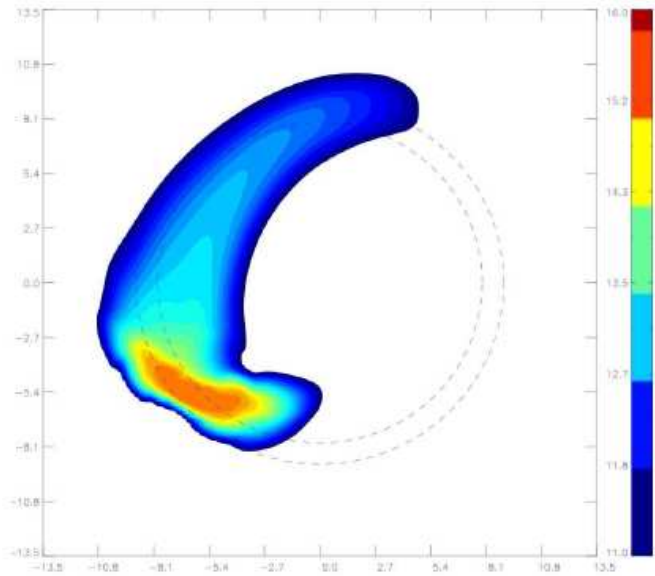

**Figure 2.** Face-on view of the z column density distribution of the SN ejecta after t = 160 Myr for a single fountain. The SNe explode at a radius R = 8.5 kpc, halfway between the radii R = 8 kpc and R = 9 kpc of the two circles drawn in the figure to guide the eye. Their distance represents approximately the maximum extension of the hole in the ISM carved by the fountain on the disk during its activity. At this time the hole has collapsed and disappeared. The column density scale is given in $cm^{-2}$ and the x - y scale is in kpc.

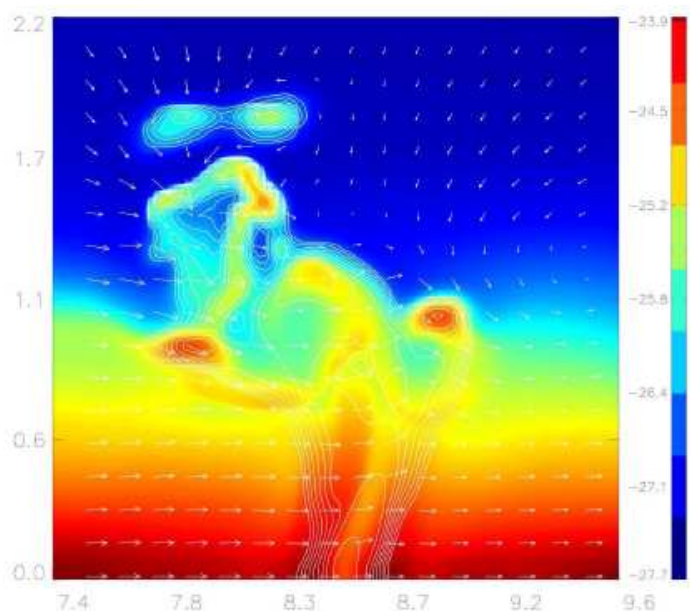

**Figure 3**. Edge-on view of the single fountain at t= 50 Myr when the ascending gas reaches its maximum height (z = 2 kpc). Isodensity curves of the ejecta are overimposed to the disk gas distribution, highlighting the fountain pattern and the cloud formation (see text).

A further insight on the evolution of the galactic fountain is obtained from Fig. 4 which illustrates several quantities starting from t=30 Myr, the time at which the SNe stop to explode. The upper panel shows how quickly the ejecta looses its angular momentum because of the interaction with the gaseous halo. After 80 Myr, nearly 10% of the angular momentum of the fountain has been transferred to the hot halo; later on, no further transfer occurs because nearly 75% of the ejecta is located below the disk-halo transition and rotates together with the ISM. The middle panel of Fig. 4 shows the ejecta mass fraction located above different heights. It is interesting to note that for t >80 Myr, the long-term evolution of these quantities becomes very slow. This is due to the fact that most of the ejecta situated above the plane is rather diluted and tends to float together with the extra-planar ISM. Finally, in the lower panel of Fig. 4, we quantify the tendency of the ejecta to move radially by plotting the fraction of the total mass of ejecta located at R >9 kpc and R <8 kpc (i.e. the radii of the two circles drawn in Fig. 2); the amount of mass located within the region 8 < R < 9 kpc is not taken into account. In the beginning, the ejecta starts to follow the expected tendency to move outward but, as the loss of angular momentum proceeds, the fraction of gas moving inward increases and after 60 Myr overrides that directed outward.

The gas lifted up by the fountain has a mass $2.5 \times 10^5$ solar masses, almost all (92%) condensed in dense filaments cooled to T = $10^4$ K (see Fig. 3). The clouds form via thermal instabilities at z ~ 2 kpc, the maximum height that the ascending gas reaches before starting to move back toward the disk at 50 Myr. The clouds have all negative z-velocities in the range 50-100 km/s. The chemical composition of these clouds is practically unaffected by the SN ejecta. The fountain is powered by 100 SNe, half of them exploding in the half space mapped by the grid; as each supernova delivers on average 3 solar masses of metals, a total mass of 150 solar masses of heavy elements is ejected by the fountain in Figs. 2 and 3. Only 20% of the metals of the ejecta is locked in the clouds, 50% of this material remains trapped within the disk and 30% remains floating over the disk as hot, diffuse gas. As a result, the metallicity increment in the clouds due to the freshly delivered metals corresponds to 0.01 in solar units and is negligible compared to the solar abundance of the ISM. As we will show below, this result also holds in models with multiple SNe associations. In conclusion, almost all the gas lifted up by the fountain condenses into clouds without being chemically affected. After 150 Myr, 45% of the fresh metals stays on the disk (below z = 800 pc) within a radial distance of R = 0.5 kpc from the OB association. A further fraction of 35% is found on the disk within the range 9.5< R <7 kpc. The remaining 20% of metals is still over the disk, half of which at R > 8.5 kpc and half at R < 8.5 kpc.

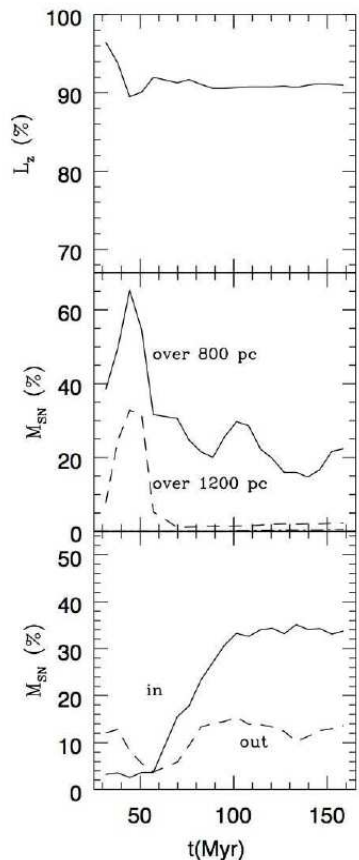

**Figure 4.** The upper panel shows the evolution of the angular momentum of the SN ejecta for the single galactic fountain of Figs. 2 and 3. The evolution of the amount of the ejecta above different heights is illustrated in the middle panel. The lower panel displays the temporal behavior of the amount of ejecta located at R < 8 kpc (dashed line) and R >9 kpc (solid line); these two radii correspond to the circles visible in Fig.2.

Figure 5 shows the results for a multiple fountain model that was produced by randomly clustered explosions of SNe originating in stellar associations spread over a disk area of 8 kpc$^2$ for a period of P=200 Myr with a mean rate adequately scaled from the rate of 1.4 x 10$^{-2}$ yr for the whole Galaxy (Cappellaro et al. 1997). During the time P of the simulation, $N_{tot}$ = 1.59 x 10$^4$ SNe exploded in the active area of the Galactic disk. The employed size-frequency distribution of the clustered SNe progenitors follows the distribution inferred from observations by Higdon & Lingenfelter (2005) for the MW.

We may notice the holes dug by the SNe into the disk in the beginning, at t=30 Myr. As time goes by, as in the case of the single fountain (of Figs. 2 and 3), the ascending material in the multiple fountains is stretched by the rotation forming comet-tail like structures. Again, as the ascending material returns to the disk it tends to fall towards the inner disk region due to angular momentum losses (of about 15%) to the halo. In this case, the maximum height attained by the fountain material, with the formation of dense cold clouds from the condensation of the hot gas, is twice larger than in the case of the single fountain and the clouds rain back to the disk with velocities between -50 km/s and -100 km/s. Thus, these results indicate that the galactic fountains are able to produce only IVCs. We have also found that these results are relatively insensitive to the distance of the fountains to the galactic centre (which were also simulated at a distance of 4.5 kpc from the galactic centre). Our simulations of multiple fountains have also revealed that the amount of gas circulating between the halo and the disk reaches a dynamical equilibrium around 150 Myr, so that after this time the amount of gas falling back on the disk is approximately equal to the ascending amount in the fountain. As before, the spreading of the SNe ejecta back in the disk is not very large. Most of the gas lifted up by the fountain falls back within a distance $\Delta R=\pm 0.5$ kpc from the place where the fountain was originated. This is in agreement with recent chemical models of the metal distribution in the MW disk (Cescutti et al. 2007).

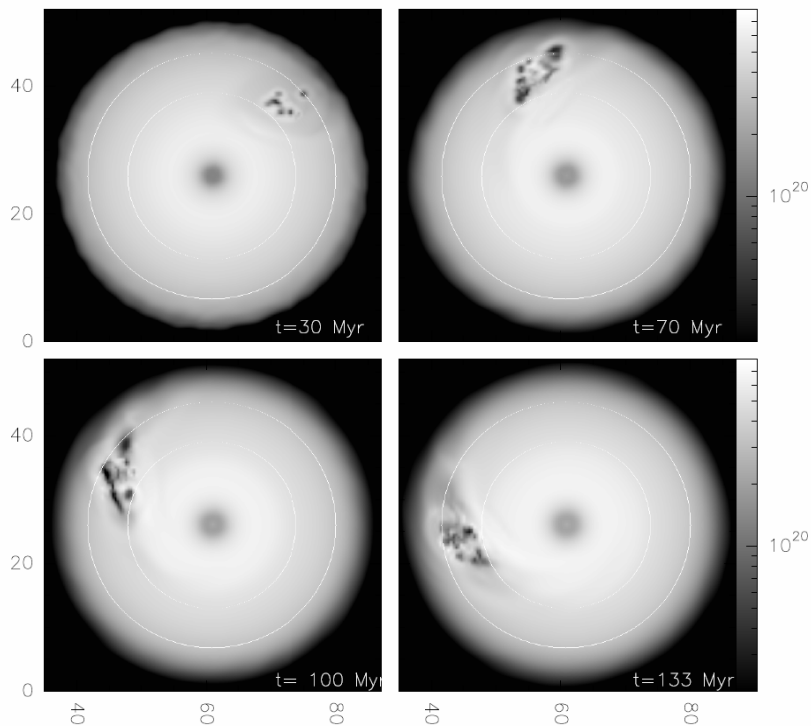

**Figure 5.** Face-on view of the evolution of a multiple fountain triggered by the explosion of SNe from randomly distributed stellar clusters over an area of 8 kpc**$^2$** of the galactic disk at a distance R=8.5 kpc from the galactic centre. The figure shows 4 snapshots of the evolution of the column density (in cm$^{-2}$) of the ascending gas (ISM+SNe ejecta) in a multiple fountain.

## 3. Conclusions and Final Remarks

Large scale 3D hydrodynamical, non-equilibrium radiative cooling simulations of random explosions of SNe in off-center stellar clusters in a rotating galactic disk-bulge system evidence the formation of giant superbubbles that break through the galactic disk into the halo forming chimneys and fountains:

- The gas lifted up by multiple fountains condenses into clouds that reach altitudes up to around 4 kpc and then fall back into the disk with velocities between -50 km/s and -100 km/s approximately. These clouds can thus explain the formation of the IVCs, but not of the HVCs that are observed at heights as large as 12 kpc in the halo. After about 150 Myr, the gas circulation between the halo and the disk in the fountains is found to reach a steady state regime and this is relatively insensitive to the galactic-centric distance where the fountains are produced.

- After a maximum lift, the clouds rain back into the disk, but towards smaller radii due to angular momentum losses (10%-15%) to the halo. This amount of angular momentum may be large enough to provoke an eventual co-rotation of the halo with the disk. However, the observed halos of star forming galaxies often show a rotational velocity gradient with respect to the disk of $\Delta v_{rot}$ = -15 km/s/kpc (for 1.3<z<5.2 kpc; Dettmar & Soida 2005). This could be an indication that some other mechanism might be operating to inhibit the halo rotation.

- The galaxy rotation inhibits both a straight vertical expansion of the fountains and the spreading of the metals transported by the fountains from the SNe. Contrary to ballistic models, most of the gas that is lifted up by the fountains falls back on the disk within a distance $\Delta R = \pm 0.5$ kpc from the place where the fountain originated. As a consequence, nearly 60% of the metals delivered by the SNe remains in the area where the fountain was formed. This small radial displacement of metals in the disk is in agreement with recent chemical models of the Milky Way (Cescutti et al. 2007).

- We have also found from the simulations that the metal pollution of the clouds is ~0.01 in solar units, therefore negligible compared to the ISM typical abundance of metals. This implies that the clouds that are formed by the fountains in the halo are almost not chemically affected by the SNe.

- Once the material of the clouds rains back into the disk we expect that it will provide the formation of new molecular clouds and supersonic filamentary structures that will feed the ISM turbulence thus closing the gas cycle between the disk and the halo (e.g., de Avillez & Breitschwerdt 2005).

- We have also performed hybrid simulations including both the ejected material from fountains triggered by SNe arising from the galactic disk and the infall of gas accreted from the IGM at a rate dM/dt = 1 to 2 solar masses per year (Melioli et al. 2008b). In this case, the formation of low-metallicity, high-velocity halo structures falling on the disk from the highest latitudes (like the HVCs) is a straightforward consequence.

A final remark is in order. In the study reviewed here, we have included all the essential ingredients of a star-forming disk galaxy, but the magnetic fields and the thermal conduction effects. The magnetic fields may be significant at helping lifting the fountains and form, e.g., at least part of the population of HVCs. In fact, large scale magnetic fields with coherent scales of

~1kpc have been observed in the halos of several of these galaxies (e.g., Dettmar 2005). The MHD simulations of galactic fountains by de Avillez & Breidtschwerdt (2005) have produced halo magnetic fields but with smaller scales. MHD effects have been also invoked by de Gouveia Dal Pino & Medina-Tanco (1999) to accelerate SNe-triggered winds in SB galaxies by a magnetocentrifugal process and more recently, Otmianowska-Mazur et al. (2007) have investigated the joint action of both the Parker-Rayleigh-Taylor instability in a galactic dynamo and the production of cosmic rays by the SNe to explain the ascension of large scales magnetic fields to the halo. These effects could also provide guidance and acceleration of the fountains and help the formation of clouds at higher latitudes than those allowed by the pure hydrodynamical simulations. Concerning the effects of the thermal conduction, these can be highly important at suppressing thermal instabilities and thus structure formation, particularly at the smallest scales (not addressed in the present study), but it may be on the other hand, inhibited by stochastic magnetic fields. These important ingredients will be considered in future work.